\documentclass[iop,numberedappendix]{emulateapj}
\usepackage{amsmath}
\usepackage[colorlinks=true,citecolor=blue]{hyperref}

\usepackage{ulem}
\usepackage{color}

\newbox\grsign \setbox\grsign=\hbox{$>$} \newdimen\grdimen \grdimen=\ht\grsign
\newbox\simlessbox \newbox\simgreatbox \newbox\simpropbox
\setbox\simgreatbox=\hbox{\raise.5ex\hbox{$>$}\llap
  {\lower.5ex\hbox{$\sim$}}}\ht1=\grdimen\dp1=0pt
\setbox\simlessbox=\hbox{\raise.5ex\hbox{$<$}\llap
  {\lower.5ex\hbox{$\sim$}}}\ht2=\grdimen\dp2=0pt
\setbox\simpropbox=\hbox{\raise.5ex\hbox{$\propto$}\llap
  {\lower.5ex\hbox{$\sim$}}}\ht2=\grdimen\dp2=0pt

\newcommand{\be}{\begin{equation}}
\newcommand{\ee}{\end{equation}}

\def\lsh{l_{\rm sh}}
\def\tdelay{\delta t}

\shorttitle{Blast wave emergence at the photosphere}
\shortauthors{Lundman \& Beloborodov}

\begin{document}

\title{A first-principle simulation of blast wave emergence at the photosphere of a neutron star merger}
\author{Christoffer Lundman\altaffilmark{1,2} and Andrei M. Beloborodov\altaffilmark{3,4}}
\affil{$^1$Department of Astronomy, Stockholm University, AlbaNova, SE-106 91 Stockholm, Sweden \\
$^2$The Oskar Klein Centre for Cosmoparticle Physics, AlbaNova, SE-106 91 Stockholm, Sweden \\
$^3$Physics Department and Columbia Astrophysics Laboratory, Columbia University, 538  West 120th Street New York, NY 10027\\
$^4$Max Planck Institute for Astrophysics, Karl-Schwarzschild-Str. 1, D-85741, Garching, Germany}

\begin{abstract}
We present the first {\it ab initio} simulation of a radiation-mediated shock emerging at the photosphere of a relativistic outflow. The simulation is performed using our code \texttt{radshock} that follows fluid dynamics coupled to time-dependent radiative transfer, calculated with the Monte-Carlo method. We use the code to examine the radiative blast wave emerging from neutron star merger GW~170817. It was previously proposed that the merger ejected a dark, relativistically expanding, homologous envelope, and then an explosion inside the envelope produced the observed gamma-ray burst GRB~170817A. Our simulation demonstrates how the shock wave generates radiation as it propagates through the envelope, approaches its photosphere, releases the radiation, and collapses, splitting into two collisionless shocks of a microscopic thickness. We find the light curve and the spectral evolution of the produced gamma-ray burst; both are similar to the observed GRB~170817A.
\end{abstract}



\section{Introduction}
\label{sec:introduction}

The breakout of a radiation-mediated shock at the photosphere of a relativistic explosion is a long-standing theoretical problem (see \citealt{WaxKat:2017} for a review). It involves time-dependent radiative transfer in the outflow, which is further complicated by possible copious production of $e^\pm$ pairs. Analytical shock breakout models have been proposed for the ``photon-poor'' \citep{NakSar:2012} and ``photon-rich'' \citep{LevNak:2020} regimes. However, direct numerical experiments have been limited to plane-parallel shocks \citep{Bel:2017, LunBelVur:2018, ItoEtAl:2018}, which describe the quasi-steady structure of shocks deep below the photosphere. The photospheric breakout is a more challenging problem, because it is time-dependent and involves a drastic transformation of the shock. The decrease in optical depth (due to spherical expansion) eventually leads to the radiation decoupling from the plasma, so the shock ceases to be radiation mediated, and this transition shapes the released radiation spectrum.

Solving the shock breakout problem became particularly important after the detection of neutron star merger GW~170817, which was accompanied by gamma-ray burst GRB~170817A. A plausible mechanism for the GRB is the breakout of a radiation-mediated shock from a cloud surrounding the merger \citep{GotEtAl:2018, BroEtAl:2018, BelLunLev:2020}. In our previous paper (\citealt{BelLunLev:2020}, hereafter BLL20) we suggested that neutron star mergers promptly eject a dark, ultra-relativistic, homologously expanding envelope, loaded with $\sim 10^4$ photons per baryon. The four-velocity of this low-mass envelope diverges toward its outer edge, and is described by a power law,

\be
u \equiv \gamma\beta \approx \left(\frac{m}{m_1}\right)^{-1/4} \qquad {\rm (envelope)},
\label{eq:speed_profile}
\ee

\noindent where $m$ is the Lagrangian mass-coordinate growing inward ($m=0$ corresponds to the outer edge), $\beta$ is the expansion speed in units of the speed of light $c$, and $\gamma=(1-\beta^2)^{-1/2}$ is the expansion Lorentz factor. The theoretically expected $m_1\sim 10^{27}$~g implies a significant scattering opacity of the relativistically outflowing plasma, which inflates the photospheric radius with time. When the central engine launches a powerful jet, it drives a blast wave into the envelope, chasing its outer layers and eventually catching up with the envelope photosphere. Viewed at an angle $\theta=20$-30$^\circ$ from the merger rotational axis, the photospheric shock emergence occurs at radius $r \sim 10^{12}$~cm. It can produce a GRB with the observed luminosity $L \lesssim 10^{47}$~erg/s and the average photon energy $\bar{E} \sim 100$~keV (BLL20), providing a promising scenario for GRB~170817A.

In this Letter we report the results of the first {\it ab initio} simulation of a photospheric shock breakout in a relativistically expanding medium. The simulation follows the propagation of a radiation-mediated shock (RMS) in the homologous envelope described by Equation~(\ref{eq:speed_profile}), with 

\begin{equation}
m_1 = 3\times 10^{27} {\rm ~g},
\end{equation}

\noindent and the photon-to-baryon ratio $n_\gamma/n_b = 10^4$ (BLL20). The homologous density profile $\rho(r,t)$ is determined by the speed profile of the envelope; its  pressure is negligible ahead of the blast wave, and so the envelope expands ballistically. The envelope has an electron-proton composition, which changes if the shock produces $e^\pm$ pairs.

The blast wave is expected to have an anisotropic power, so that its Lorentz factor $\Gamma$ can vary with the polar angle $\theta$. However, the blast wave is casually disconnected on angular scales $\delta\theta > \Gamma^{-1}$. Therefore, we approximate its dynamics at given $\theta$ as part of a spherically symmetric explosion with isotropic energy equivalent ${\cal E} = 4\pi (d{\cal E}/d\Omega)$. The only free parameter of the blast wave is its initial energy per unit solid angle $d{\cal E}_0/d\Omega$. We choose it so that it gives a GRB with isotropic equivalent ${\cal E}_\gamma \sim 10^{47}$~erg, consistent with GRB~170817A.\footnote{The shock energy ${\cal E}$ decreases from its initial values ${\cal E}_0$ as it propagates in the homologous envelope, whose density decreases with time and radius. This is a known property of type II self-similar solutions in hydrodynamics (Zeldovich \& Raizer 1966).}

We wish to find: (1) how the RMS transforms at the photosphere and (2) the light curve and spectral evolution of the produced GRB, which can be compared with observations. We use an extended version of \texttt{radshock}, a radiation hydrodynamics code designed specifically for RMS simulations \citep{LunBelVur:2018}. Here, \texttt{radshock} couples Lagrangian hydrodynamics to Monte-Carlo radiative transfer in spherical coordinates. The code uses the full Klein-Nishina cross section for scattering and self-consistently computes photon-photon pair ($e^\pm$) production; it therefore automatically follows the process of photon Comptonization and (if high-energy photons are generated) pair production inside the RMS. Other technical advances include variable mass binning and the evaluation of hydrodynamical source terms from the moments of the radiation field. For the simulation presented below, we used $24000$ hydrodynamical mass bins, and $10^7$ Monte-Carlo photons.

The code calculates the blast wave emission from first principles, with one caveat: collisionless shocks, which develop outside the photosphere, are not resolved to the microscopic plasma scale. Instead, they are treated as hydrodynamic discontinuities, which satisfy the shock jump conditions. This description correctly results in nearly impulsive heating of electrons and protons in the collisionless shock. The electrons become relativistically hot and quickly cool on the overlapping radiation field, decoupling thermally from the protons. The code follows this decoupling and the subsequent Coulomb energy exchange between electrons and protons. What the code cannot follow is nonthermal particle acceleration, which would require a kinetic plasma simulation. However, our result demonstrates that the energy budget available for particle acceleration in the RMS breakout is negligible, and so there is no need for a kinetic simulation.


\section{Photospheric emergence of the RMS}

\subsection{Blast wave in the optically thick envelope}
\label{sec:blast_wave}

The photospheric shock emergence occurs at the Lagrangian mass coordinate $m_\star\sim 10^{26}\,$g where the envelope has $u(m_\star)\approx 3$ (BLL20). The blast wave dynamics in deeper layers $m\gg m_\star$ follows a hydrodynamic solution with the adiabatic index of $4/3$. To save computational time, our simulation begins to follow the blast wave after it has reached $m_i=7 \times 10^{26}\,$g and continues its expansion toward $m_\star$. The initial shock location is still deep below the photosphere, at the Thomson optical depth of 900. Note that $u(m_i)\approx 1.4$ (Equation~\ref{eq:speed_profile}). The simulation tracks the photons that are initially contained in the envelope at $0 < m < m_i$. This range covers the region where the GRB spectrum and light curve will form.

The blast wave is launched by smoothly accelerating a piston at the inner boundary $m_i$ to four-velocity $u_p = 3.25$. We chose $u_p$ so that it gives the observed GRB energy $\lesssim 10^{47}\,$erg. One  must also choose the time of launching the blast wave $t_i$ or, equivalently, the radius of shell $m_i$ at the beginning of the simulation, $r_i\approx v(m_i)t_i\approx 0.8ct_i$. This choice is related to the explosion lag behind the gravitational waves, which propagate at $r_o=ct_i$. We chose $t_i=6\,$s; it will lead to the observed GRB delay of $\sim 1.5\,$s, close to the delay of GRB~170817A.

The shock propagating in the optically thick envelope is mediated by photon scattering. Like any RMS, it has the width $\lsh \approx (c/v_u)\lambda$ \citep{BlaPay:1981b}, where $v_u$ is the upstream speed, $\lambda$ is the photon mean free path, and all the quantities are measured in the downstream rest frame. The plasma crosses the shock on the timescale

\be
\Delta t_{\rm sh}=\frac{\lsh}{v_u}\approx \frac{t_{\rm exp}}{\beta_u^2\tau},
\ee

\noindent where $\beta_u=v_u/c$, $t_{\rm exp}\approx r/c\Gamma$ is the shock expansion timescale, and $\tau=ct_{\rm exp}/\lambda$ is the characteristic optical depth ($\tau>1$ below the photosphere). The RMS is quasi-steady as long as $\Delta t_{\rm sh}\ll t_{\rm exp}$. This condition is satisfied deep below the envelope photosphere. Here the shock structure would be reproduced by a plane-parallel simulation; its radiation spectrum is similar to that found in \citet{Bel:2017} and \citet{LunBelVur:2018}.

When viewed in the lab frame, the upstream has the Lorentz factor $\gamma$ (Equation~\ref{eq:speed_profile}) and the downstream moves with $\Gamma>\gamma$. The rising profile of the envelope velocity (Equation~\ref{eq:speed_profile}) implies that the blast wave must accelerate to $\Gamma \gg 1$. It chases the outer layers of the envelope with ever increasing $\gamma$, however the ratio $\Gamma/\gamma < 2$ weakly changes, so that the shock amplitude remains mildly relativistic (BLL20). Coincidentally, in the simulation presented in this {\it Letter} the RMS amplitude remains just below the threshold for copious $e^\pm$ production $\gamma_u\beta_u \approx 1$ \citep{LunBelVur:2018}.

The RMS converts the upstream kinetic energy of protons $(\gamma_u-1)m_p c^2$ to heat dominated by radiation (whose initial upstream enthalpy is negligible). The resulting radiation energy density $U_\gamma$, measured in the downstream rest frame, is related to the mass density $\rho$ by

\be
\frac{U_\gamma}{\rho c^2}\approx \gamma_u-1 \sim 0.1.
\ee

\noindent The plasma temperature profile across the RMS closely tracks the local Compton temperature\footnote{The Compton temperature for a given radiation field is defined as the electron temperature for which the net energy exchange between electrons and radiation through scattering vanishes.} of radiation $T_{\rm C}$, which rises behind the shock, because of the dissipation. Its exact value is not far from the average photon energy,

\be 
kT_{\rm C}\sim \frac{U_\gamma}{n_\gamma} = m_pc^2 \frac{n_p}{n_\gamma} \frac{U_\gamma}{\rho c^2} \sim 10 \, \mathrm{keV},
\ee

\noindent where $n_\gamma/n_p = 10^4$ is the photon number per proton in the homologous envelope and $k$ is the Boltzmann constant. We find that photon production by double Compton scattering and bremsstrahlung from the heated plasma behind the shock do not significantly change the photon number, so that $n_\gamma/n_p$ stays constant. The magnetic field in the homologous envelope is expected to be very weak (BLL20). Therefore, the RMS does not have any significant collisionless subshock that would form in a magnetized plasma \citep{Bel:2017} and produce synchrotron photons \citep{LunBel:2019}.

The proton and electron temperatures $T_p\approx T_e$ remain coupled by Coulomb collisions, so both track the Compton temperature $T_{\rm C}$. The Compton equilibrium is enforced on the timescale $t_{\rm C}=3 m_e c/4U_\gamma \sigma_\mathrm{T}$ (measured in the fluid rest frame), which is much shorter than the plasma crossing time of the shock,

\be
\frac{t_\mathrm{C}}{\Delta t_{\rm sh}}\sim Z_\pm\,\frac{m_e}{m_p}\,\frac{\rho c^2}{U_\gamma}\,\beta_u^2\ll 1,
\label{eq:t_c}
\ee

\noindent where $Z_\pm = n_e/n_p$ is the number of electrons and positrons per proton; the simulation gives $Z_\pm \approx 1$ as $e^\pm$ pair creation turns out inefficient. Note that $t_{\rm C}/\Delta t_{\rm sh}$ is independent of the optical depth $\tau > 1$. The protons and electrons exchange energy via Coulomb collisions on the timescale \citep{Stepney:1983},

\be
t_{ep} = \sqrt{\frac{\pi}{2}}\frac{m_p}{m_e}\frac{1}{\sigma_\mathrm{T} n_e c \ln\Lambda} \left(\frac{kT_e}{m_ec^2}  + \frac{kT_p}{m_pc^2} \right)^{3/2},
\label{eq:t_ep}
\ee

\noindent 
where $\ln\Lambda \approx 15$. At the subphotospheric stage of the RMS propagation $T_p \approx T_e$  is ensured by $t_{ep} < \Delta t_{\rm sh}$.

\begin{figure}
\includegraphics[width=\linewidth]{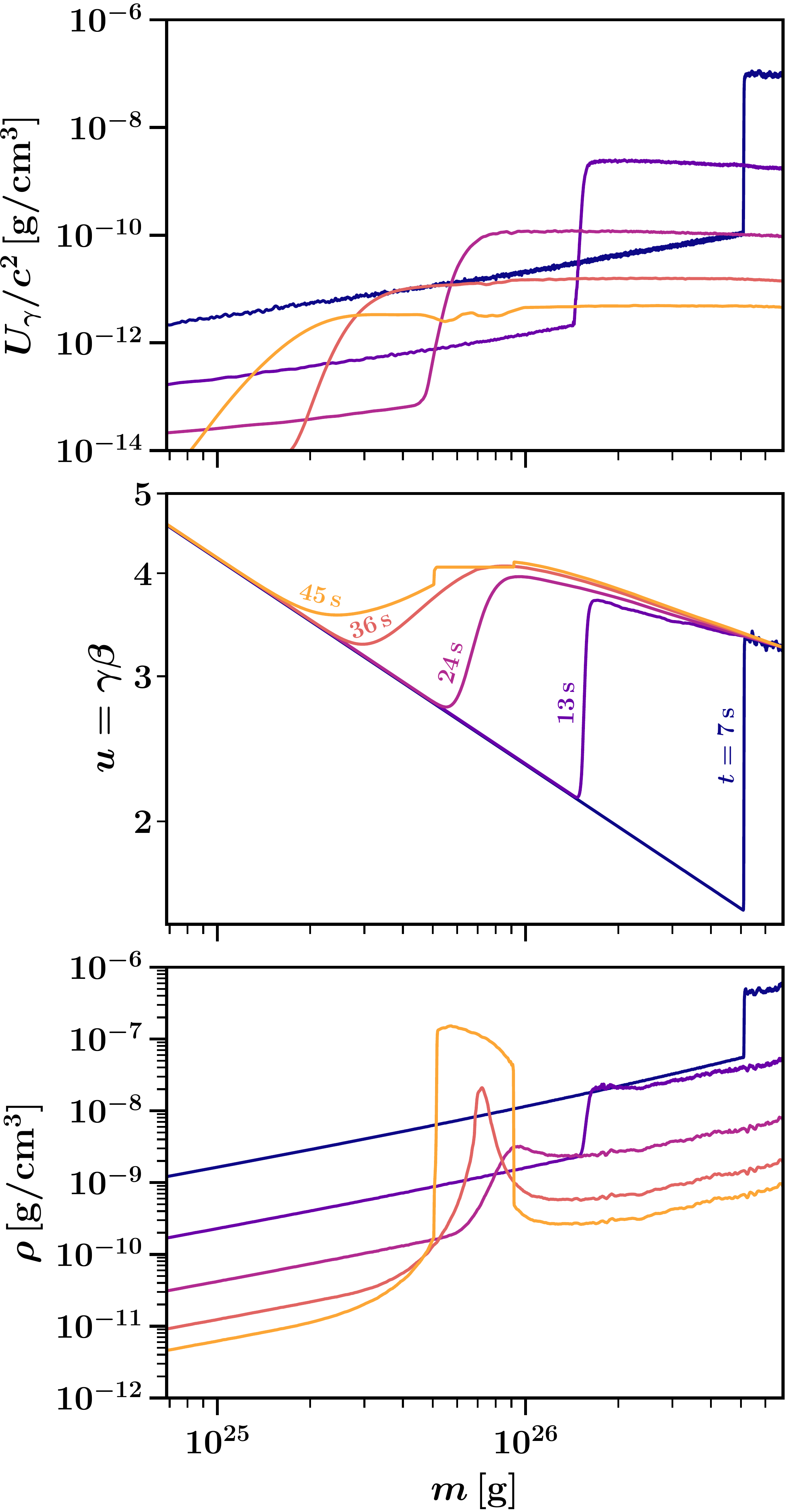}
\caption{Photospheric emergence of the RMS. The proper radiation energy density $U_\gamma$, plasma four-velocity $u = \gamma\beta$ (measured in the fixed lab frame), and proper mass density $\rho$ are shown as functions of the Lagrangian mass-coordinate $m$. The mass $m$ is measured from the outside, so that $m=0$ at the outer edge of the envelope. The $U_\gamma(m)$, $u(m)$, and $\rho(m)$ are shown at five times  $t = {23, 32, 43, 58, 79}$~s, with five colors indicated in the figure. Here the times are measured in the fixed lab frame. The approximate radial position of the shock at time $t$ is $r_{\rm sh}(t)\approx r_i+c(t-t_i)$.}
\label{fig:plasma_evolution}
\end{figure}

The simulation shows that the RMS stays in this standard, steady state until it approaches the photospheric radius $r_\star\sim 10^{12}\,$cm at the lab-frame time $t_\star\sim 20\,$s. Before this time, the shock expansion follows the hydrodynamic solution with adiabatic index $\gamma_{\rm ad} = 4/3$, and its narrow front structure $\Delta r_{\rm sh}\ll r_{\rm sh}$ (where the photons are Comptonized) is close to the plane-parallel model.

\subsection{RMS approach to the photosphere}

Clearly, the RMS is unsustainable outside the photosphere, because the photon mean free path will exceed $r$, and the RMS thickness would exceed the size of the system. Most of the RMS energy must be radiated away, and part of it will be carried by plasma motions, which can steepen into collisionless shocks. Figure~\ref{fig:plasma_evolution} shows five snapshots from the simulation, demonstrating the RMS transformation at $r\sim r_\star$.

At lab-frame time $t = 7\,$s the RMS is still below $r_\star$ and near the end of the plane-parallel phase. In the later snapshots, the plasma four-velocity profile across the RMS, $u(m)$, becomes increasingly broad and shallow. This is caused by the gradient of $\gamma(m)$ in the envelope. Deep below the photosphere, the shock thickness in the Lagrangian mass-coordinate $m$ satisfies $\Delta m_{\rm sh} \ll m$, so the gradient of $\gamma(m)$ in the upstream medium only weakly affected the shock front structure $u(m)$. When the RMS approaches the photosphere, its thickness grows to $\Delta m_{\rm sh}/m \sim 1$; now the gradient of $\gamma(m)$ significantly reduces the RMS velocity jump.

The decrease in scattering opacity near the photosphere weakens the coupling of photons to the plasma. As a result, the dense radiation begins to leak out from the downstream into the upstream---a large $U_\gamma$ develops ahead of the maximum of $u(m)$. This is seen to occur beginning from the $t=24\,$s snapshot.

Furthermore, the radiation loses its grip on the plasma. Radiation pressure is responsible for the propagation of any RMS, and the smoothening of the radiation pressure gradient across the RMS implies that the shock ``stalls:'' the profile of $u$ slows down its propagation in the Lagrangian mass-coordinate $m$. In particular, the position of the maximum of $u(m)$ stalls at $m \approx 10^{26}$~g, as can be seen by comparing the $U_\gamma$ and $u$ panels of Figure~\ref{fig:plasma_evolution}. The plasma motion gradually transitions toward a quasi-ballistic regime, which inherits the broad RMS velocity profile.

The short Compton timescale $t_\mathrm{C}$ (Equation~\ref{eq:t_c}) implies that radiation continues to control the plasma temperature also after the radiation begins to decouple from the plasma: $T_p\approx T_e\approx T_{\rm C}$. The temperature is low, and the plasma flow is highly supersonic.

\subsection{Caustic in the plasma flow: launching two collisionless shocks}

The supersonic plasma flow has a gradient $dv/dm > 0$ inherited from the RMS velocity profile, which corresponds to $dv/dr < 0$. This leads to the development of a caustic in the quasi-ballistic plasma flow. The caustic develops at $m \lesssim 10^{26}$~g, where $|dv/dr|$ is maximum, slightly ahead of the maximum of $u(m)$. The converging motion with $|dv/dr|_{\max}$ enhances the local plasma density $\rho$ on the timescale $|dv/dr|_{\max}^{-1}$. It is shorter than the flow expansion time, and so the density spike develops quickly (compare the density profiles at $t=36\,$s and $t=45\,$s in Figure~\ref{fig:plasma_evolution}).

Before a true caustic could form (i.e. before $\rho$ diverges), the growing gas pressure $P_g\propto \rho T_{\rm C}$ stops the compression and launches two collisionless shocks in opposite directions: a forward and a reverse shock. This process is illustrated in Figure~3 in \citet{Bel:2017}. The compression stops when $P_g$ matches the ram pressure of the converging flow. This occurs when the density spike has grown by more than 3 orders of magnitude,

\be
\frac{\rho_{\rm spike}}{\rho}\sim \frac{n_\gamma}{n_e + n_p} \sim 5 \times 10^3.
\ee

\begin{figure}
\includegraphics[width=\linewidth]{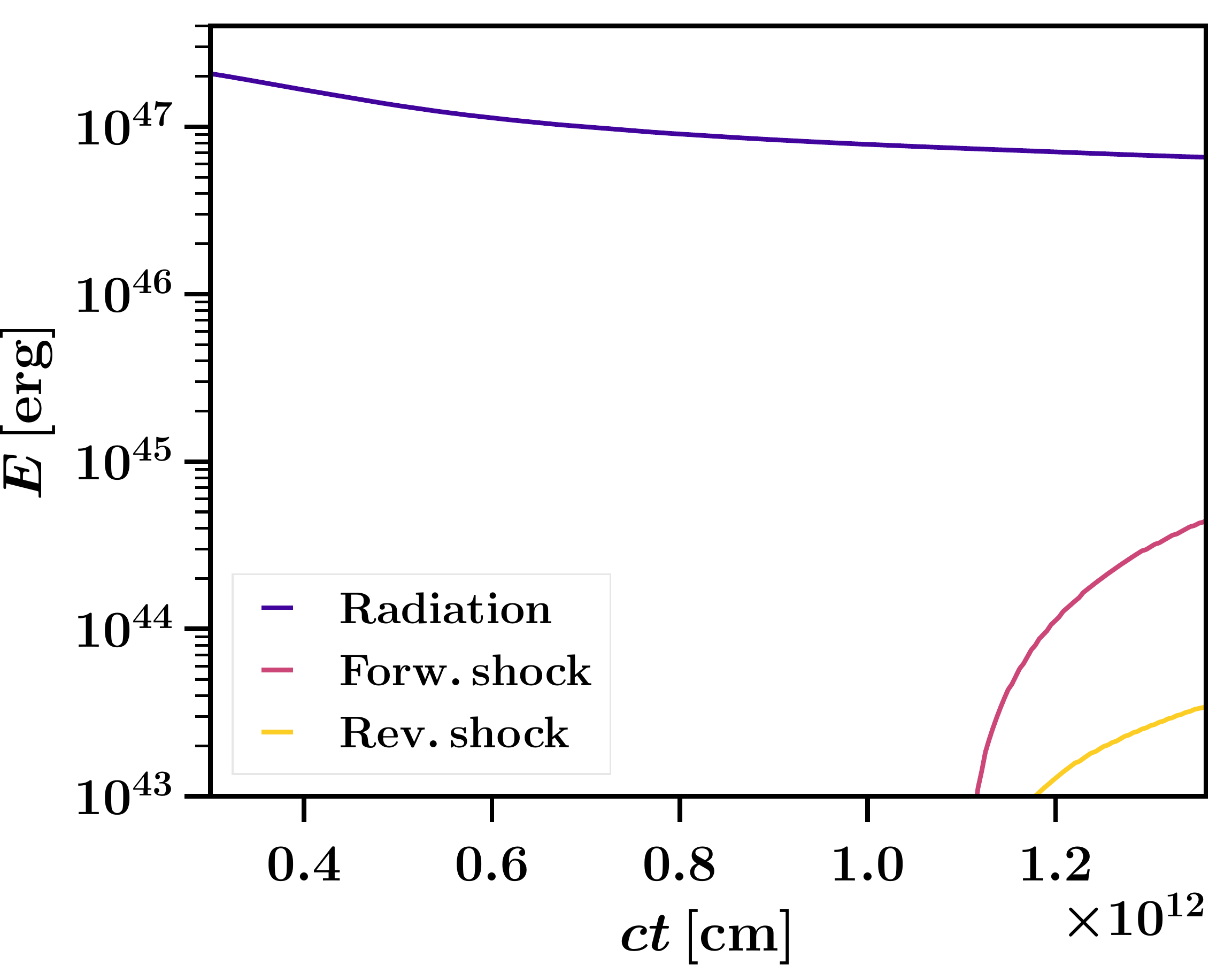}
\caption{Cumulative energy dissipated in the forward (purple) and reverse (yellow) collisionless shocks, as a function of lab-frame time $t$. The total energy carried by radiation is shown by the blue line.}
\label{fig:collisionless_energy_budget}
\end{figure}

As one can see from Figure~\ref{fig:plasma_evolution}, the forward shock is stronger than the reverse one. However, both collisionless shocks remain much weaker than the original RMS. The energy budget left for the collisionless shocks is only a few percent of the escaping radiation energy, as shown in Figure~\ref{fig:collisionless_energy_budget}. The collisionless shocks are weak because the plasma velocity profile in the RMS becomes shallow at the photosphere. Furthermore, the nascent forward shock does not have a chance to significantly strengthen at later times, as it propagates in a medium with a rising $\gamma$, which has been additionally accelerated by the escaping radiation.

The plasma is impulsively heated in the collisionless shock and cooled behind it. The electrons are cooled on the short Compton timescale $t_{\rm C}$.\footnote{The fast cooling corresponds to a short spatial scale, which is challenging to resolve. Our code \texttt{radshock} solves this problem by using non-uniform mass binning; in particular, we use the highest mass resolution around $m_\star$, in the region where the collisionless shocks form. Furthermore, the radiation and hydrodynamical timesteps are decoupled, which allows the (computationally cheaper) hydrodynamics to evolve with extremely short timesteps.} The protons cannot emit significant radiation and their cooling relies on the electron-proton coupling, which is a slower process. As a result, a layer of hot protons forms behind the collisionless shock. The proton energy is passed to the electrons, and then to radiation. This effect is followed by the simulation and found to not significantly affect the observed burst. The simulation also shows that the electron temperature behind the weak collisionless shocks does not greatly exceed $m_ec^2$. Therefore, the inverse Compton photons generated by the shocks do not cause significant creation of $e^\pm$ pairs.

A yet smaller luminosity must be produced by nonthermal particles accelerated in the collisionless shocks. The low magnetization of the envelope implies that the shocks are mediated by Weibel instability, and such shocks efficiently accelerate particles (e.g. \citealt{SirSpi:2011}). Following this process would require a plasma kinetic simulation. However, the energy budget of nonthermal particles is below 1\% of the burst energy, and their emission is negligible.


\section{The burst light curve and spectral evolution}
\label{sec:obs}

The simulation self-consistently tracks radiative transfer throughout the shock evolution to transparency. After the scattering probability for a Monte-Carlo photon becomes small, we assume that it propagates freely and determine its time of arrival to a distant observer $t_{\rm obs}$. The observer time is measured relative to the first signal arriving from the merger, i.e. relative to the arrival of the gravitational waves. The delay in launching the blast wave, and its propagation with speed slightly less than $c$, leads to the delay of the GRB.\footnote{By contrast, the homologous envelope is ejected promptly after the merger, on a timescale $t_{\rm obs}\ll \tdelay$. The outer edge of the envelope expands with speed $c$ immediately behind the gravitational waves (BLL20).} Sorting the $10^7$~Monte-Carlo photons by their arrival times and energies then gives the burst spectrum as a function of $t_{\rm obs}$. Integrating (summing) over photon energies gives the observed bolometric light curve.

\begin{figure}
\includegraphics[width=\linewidth]{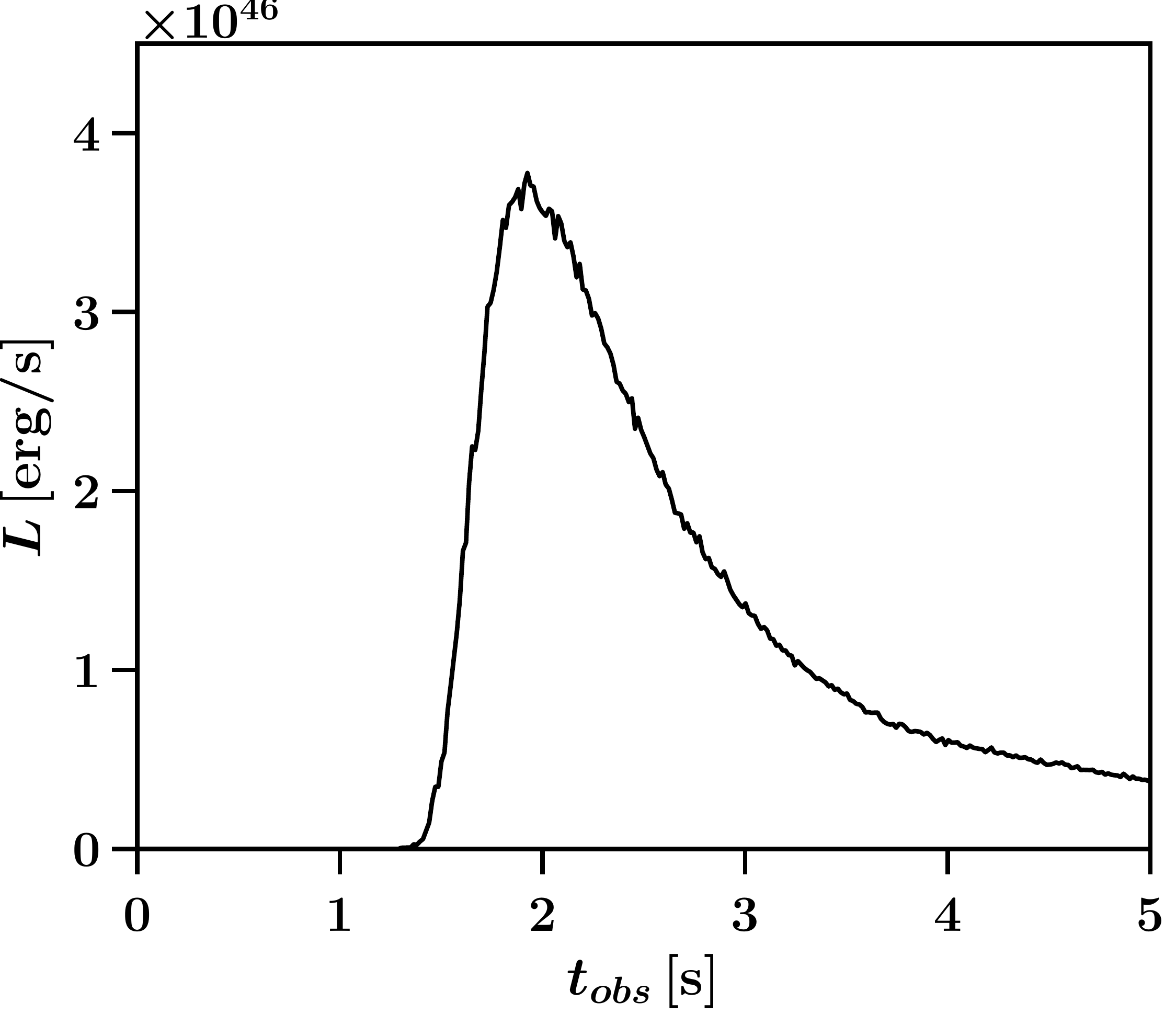}
\caption{GRB light curve predicted by the simulation ($t_{\rm obs} = 0$ corresponds to the arrival of the gravitational waves). The kink at $t_{\rm obs}\approx 4.2\,$s is artificial --- the simulation misses some emission at later times (see text).}
\label{fig:light_curve}
\end{figure}

\begin{figure}
\includegraphics[width=\linewidth]{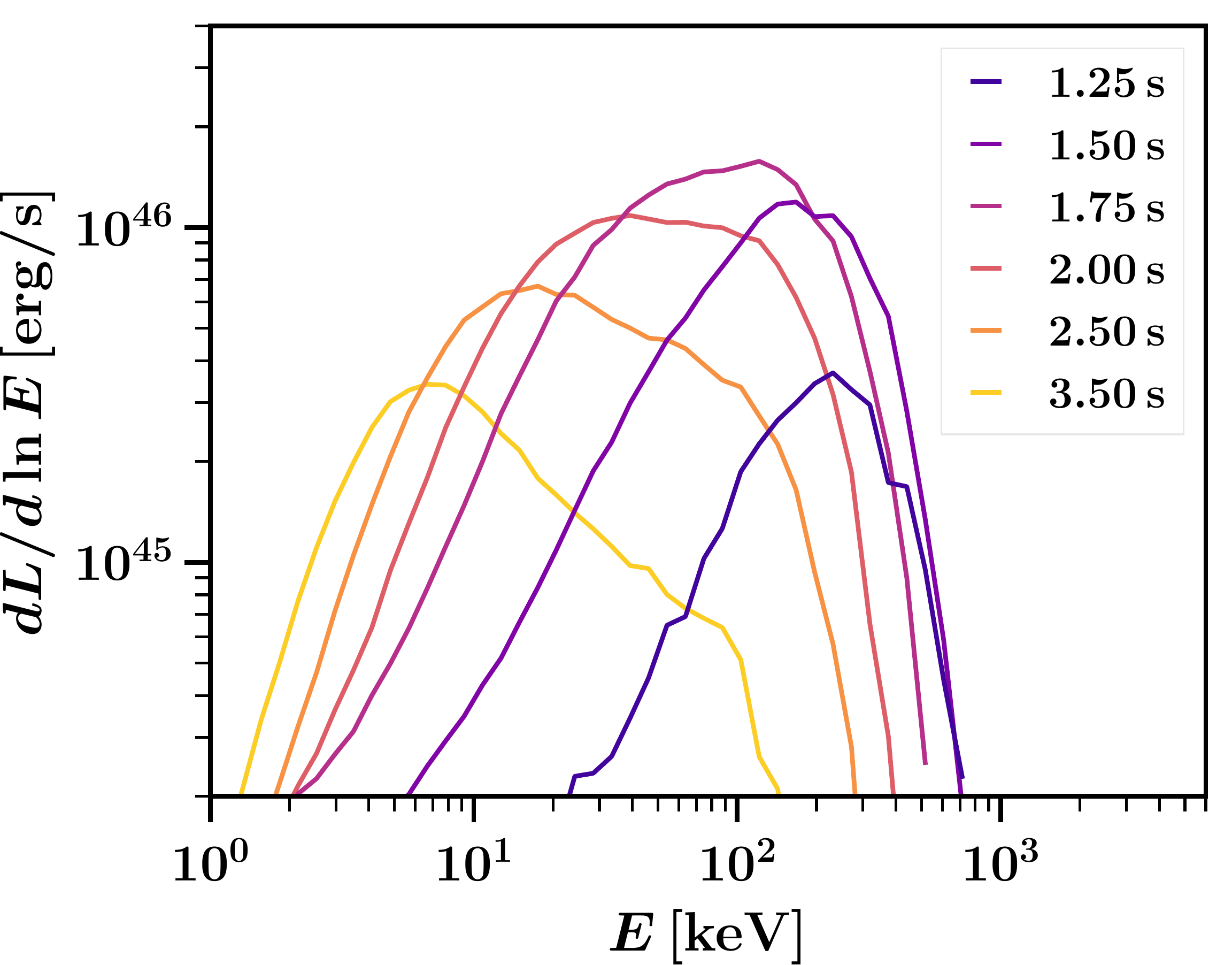}
\caption{Simulated evolution of the observed GRB spectrum. Different colors correspond to different $t_{\rm obs}$ indicated in the figure.}
\label{fig:spectral_evolution}
\end{figure}

Surprisingly, the simulation demonstrates that the radiative transfer shaping the burst is not complicated by $e^\pm$ creation (which would delay the transition to transparency) or by the collisionless shock formation. Instead, the burst is shaped simply by how the RMS radiation leaks out at the photosphere. Figure~\ref{fig:light_curve} shows the bolometric light curve  of the produced burst, and Figure~\ref{fig:spectral_evolution} shows its spectral evolution.

The earliest emission (first spectrum in Figure~\ref{fig:spectral_evolution}) is dominated by energetic photons, because they see a smaller opacity due to the Klein-Nishina correction, and because the early arrival time correlates with a nearly radial propagation direction, which gives the strongest Doppler shift to the photons. The Comptonized radiation in the middle of the RMS has a hard spectrum. This main part escapes almost immediately during the shock breakout and forms the peak of the burst. Its average photon energy $\bar{E} \sim 100$~keV is the result of $n_\gamma/n_p = 10^4$ in the expanding envelope, as discussed in BLL20. Finally, there are photons residing further downstream, behind the RMS. These photons were heated earlier, when the RMS was still below the photosphere. They are released later and hence experience more scatterings inside the expanding plasma. The scatterings cool the radiation adiabatically, and also partially thermalize its spectrum.

The simulations results are remarkably similar to GRB~170817A. The light curve has a single peak with a rise time of $\sim 0.5$~s, a width of $\sim 1$~s, and a luminosity of $\sim 4 \times 10^{46}$~erg/s. These features are in quantitative agreement with GRB~170817A \citep{GolEtAl:2017,VerEtAl:2018}. At the same time, the simulation reproduces the observed spectral evolution of GRB~170817A (see Figure~1 in \citealt{VerEtAl:2018}). The initial phase of the burst is very hard, with a typical photon energy $E \sim 300$~keV. The burst remains hard, $E \sim 100-200$~keV during the main peak and then quickly softens, so that by $t_{\rm obs} = 2.5$~s the spectrum peaks at $E \sim 20$~keV.


\acknowledgments
CL is supported by the Swedish National Space Board under grant number Dnr. 107/16. AMB is supported by NSF grant AST-1412485, a Simons Investigator Award (grant \#446228), and the Humboldt Foundatio.n

\bibliographystyle{apj}
\bibliography{ref}

\end{document}